# Alternatives for Testing of Context-Aware Contemporary Software Systems in industrial settings: Results from a Rapid review

Santiago Matalonga. School of Computing, Engineering, and Physical Science. ALMADA Research Centre. University of the West of Scotland, Paisley, Renfrewshire, United Kingdom, santiago.matalonga@uws.ac.uk

Domenico Amalfitano. Department of Electrical Engineering and Information Technology, DIETI. University of Naples Federico II, Naples, Italy, domenico.amalfitano@unina.it

Andrea Doreste. System Engineering and Computer Science, COPPE. Federal University of Rio de Janeiro, Rio de Janeiro, Brazil, doreste@cos.ufrj.br

Anna Rita Fasolino. Department of Electrical Engineering and Information Technology, DIETI. University of Naples Federico II, Naples, Italy, annarita.fasolino@unina.it

Guilherme H. Travassos. System Engineering and Computer Science, COPPE. Federal University of Rio de Janeiro, Rio de Janeiro, Brazil, ght@cos.ufrj.br

# Abstract

**Context**: Context-aware contemporary software systems (CACSS) are mainstream. Furthermore, they present challenges for current engineering practices. These challenges are distinctively present when testing CACSS, as the variation of context deepens the limitations of available software testing practices and technologies.
**Objective**: *To u*nderstand how the industry deals with the variation of context when testing CACSS.
**Method**: A Rapid Review was commissioned to uncover the necessary evidence to achieve the objectives.
**Results**: Our results show that current research initiatives aim to generate or improve Test Suites that can deal with the variation of context and the sheer volume of test input possibilities. To achieve this, they mostly rely on modeling the systems' dynamic behavior and increasing computing resources to generate test inputs. We found no evidence of research results aiming at managing context variation through the testing lifecycle process.
**Conclusions**: We discuss how the identified solutions are not ready for mainstream adoption. They are all domain-specific, and while the ideas and approaches can be reproduced in different settings, the technologies noon to be re-engineered and tailor to the specific CACSS.

# 1 Introduction

Contemporary software systems (CSS) are now mainstream. We use the term CSS to refer to software systems that demand the integration of devices and communications technologies [1] (systems that integrate IoT, Blockchain, self-driving autonomous vehicle are examples of CSS). These systems are getting attention in the press, mainly when they are context-aware. The hype of the press went from CEOs arguing that "driverless trucks will never happen" [2] to headlining the first successful coast-to-coast driverless truck journey in the USA [3]. The hype of context-aware contemporary software systems (CACSS) is well deserved, as such software systems are bound to enter several aspects of our daily lives. However, we have been arguing that CACSS pose challenges to our current engineering technologies [4]. Their testing exposes the limitations of available software testing practices and technologies. Context-awareness amplifies the software testing conundrum between coverage and effort [5], as the variation of context greatly increases the test input space. We claim that these systems are currently well accepted and deployed in safety-critical underregulated domains (like Autonomous vehicles). Therefore, they can present new forms of risk.

Not considering the context nature during testing and quality assurance is costing lives! Examples are already in the news, like the Boeing 737 Max [6] and the Chevrolet Onix's recall in Brazil [7]. In the Boeing 737 Max case, there was an omission to consider all design changes to the airplane aerodynamics when updating a software component designed to assist pilots in potentially dangerous situations. This omission led to a scenario where the software component enhanced the danger due to a faulty context interpretation. Such a scenario was (apparently) never tested before deployment [6]. In the Chevrolet Onix case, the car sensors were not prepared to detect the exceptional heatwave in Rio the Janeiro and led to situations where the car batteries ignited [7].

Based on [8], this work defines context as any piece of information that may be used to characterize an entity's situation (logical and physical objects present in the system's environment) and the relations relevant to the actor-computer interaction between actors and computers. Context-awareness is a dynamic property of a software system that can evolutionarily affect its overall behavior in the interaction between actors and computers [9]. Therefore, context-aware contemporary software systems can identify changes in the logical or physical environment (i.e., context) and adapt their behavior to provide better service to the actor. In the relationship between the CACSS and the context lies the problem with their testing. When the context and variation are considered, the input space for stimulating the CACSS grows beyond our current technology's handling capacity. The main strategy to deal with the context and its variation is to model the system and exploit computing resources in the form of simulation, as found through our previous research and experience with the testing of CACSS under the umbrella of the CAcTUS (Context-Aware Testing for Ubiquitous Systems) project. We argue [10], and so have others [11], that whether simulations and formal verifications are the most popular approaches, they are not prepared yet to deal with all the possible variations of the context surrounding the CACSS.

Under the assumption that CACSS is mainstream, we undertook a Rapid Review [12] to understand how the industry deals with the variation of context when testing CACSS. Rapid Reviews are a secondary research method aimed at uncovering evidence to help solve a practical problem. Given our working assumption, knowledge of the literature in the field, and practical experience in testing CACSS, we claim that a Rapid Review is a suitable research method to achieve our research objectives.

Our results show that the focus of current research initiatives is aimed at generating or improving Test Suites that can deal with the variation of context and the sheer volume of test input possibilities. To achieve this, they mostly rely on the modeling of the dynamic behavior of the systems. They need to rely on an increasing usage of computing resources to generate test inputs. Furthermore, we found no evidence of research results aiming at managing the variation of context throughout the testing lifecycle process. In addition to this, all identified solutions are domain-specific and are not ready for widespread transfer. It means that software tester practitioners looking to test CACSS need to take on these approaches and re-engineer -or tailor - the solutions to their working domains, which comes at a considerable cost.

The remainder of this paper is organized to convey the research method and results. First, we aim to relate state of the art in testing CACSS (section 2) and establish the concepts included in our outlook on testing these systems (section 3). Then we provide details of the design (section 4) and the research method's execution (section 5). The remaining sections are devoted to data analysis (section 6), threats to validity (section 7), and discussions (section 8).

## 2 Related Works

As we have mentioned in the introduction, we have been working with the testing of CACSS for about a decade, formally coming together about 2015 under the CAcTUS (Context-Aware Testing for Ubiquitous Systems) project umbrella. From 2015 to 2017, we looked at the aspects that made testing the context-aware feature of software systems challenging. We explored this from several points of view. In [13], we describe the execution of a *quasi*-systematic literature review to identify testing techniques to support the test of CACSS. In [9], we describe the execution of another *quasi*-systematic literature review to observe how test cases were being designed for context-aware software systems. This section describes research related to our work lines but was not carried out by our research group.

Regarding secondary studies, [14] presents a systematic mapping study that was performed to identify Android testing tools and described the suitability for testing context-aware applications. The work by [15] presents a survey of simulation techniques for testing context-aware android applications.

Regarding primary studies, we highlight the emergence of a few innovations since the CACTUS-based papers. First, cyber-physical systems (autonomous vehicles [16] ) have crept in as suitable domains to which context-aware technologies can be deployed. Regarding tools and solutions for testing CACSS, we highlight three emerging research trends. Artificial intelligence is applied to generate or derive test cases (for instance, see [17]). Search-based techniques are applied to reframing the problem of finding suitable pairs of test inputs and test outputs (for instance, see [18,19]). Finally, a technique that has been called "runtime testing" [20] or "*in-vivo* testing" [21] embodies the idea that copies (or models) of the test item can be deployed side-by-side. The strategy aims at identifying potential failures before the CACSS processes them (for instance, see [22]).

Given the current state of the art and our ongoing assumption that CACSS are mainstream, the work described in this paper looks at identifying proposals that have been experienced in an industrial setting. Furthermore, we differentiate ourselves from our previous work. We are looking at a specific context-aware software system, denouncing context-aware contemporary software systems (see section 3.1).

# 3 Terms and Definitions

In this section, the concepts and the terms used in our study are presented. More in detail next sections describe (1) definitions and examples of Context-Aware Contemporary Software Systems (CACSS), (2) challenges and issues for testing Context-Aware Software Systems, and (3) terms related to the testing processes according to the ISO/IEC/IEEE 29119.

## 3.1 Contemporary Software Systems

Contemporary software systems (CSS) exploit technologies that offer challenges for their construction since they call into question our traditional form of developing software. Usually, such software systems demand the integration of devices and communications technologies [1]. Ubiquitous systems, the Internet of Things, Smart Cities, Industry 4.0, and cyber-physical systems are typical CSS examples. In these systems, physical objects with embedded software are interconnected by networks to provide the expected behavior. Besides, sensors, computer devices, and different applications should interact, exchange information, and work with distinct elements to guarantee adequate functionalities.

Therefore, CSS are, in general, context-aware. Their behavior must conform with the context it is being used. Our previous research regarding ubiquitous [23,24], Internet of Things [1,25], and context-aware software systems [9,10,13] revealed some gaps and the need for software technologies to support the engineering of CSS [25], mainly when their non-functional requirements include context-awareness.

## 3.2 Issues and Challenges in Context-Aware Software Testing

As we have mentioned in our previous research, we highlight the many challenges and issues that must be properly addressed when testing context-aware software systems [13]. In that research, we uncovered evidence that processes that have been usually adopted for testing traditional software systems are adapted for considering the peculiarities of context-aware software systems. They tend to fix the value of context variables during the test case design or execution. Fixing the context variable's value is the most used approach to control both the context-dependent test input and output, as depicted in Figure 1. Such techniques suffer from the following limitations:

1. they cannot completely cover all the test input space of a context-aware application. Therefore, this solution limits the possible coverage of the resulting test cases.

2. it is unlikely that a test oracle can be defined for all possible values (and/or a combination of) values that can stimulate the context-aware test item.
3. it is not feasible to define a test oracle for each possible combination of context variable values.

However, we claim that these techniques do not consider the variation of context during testing execution.

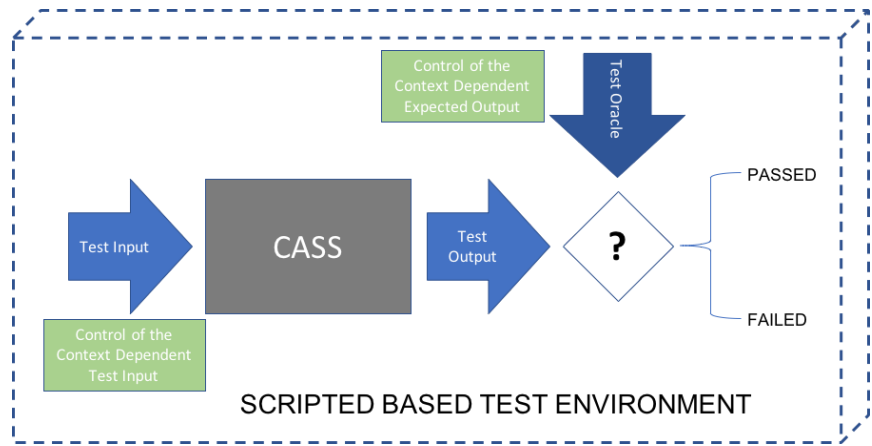

Figure 1. Traditional Testing Process adapted for Context-Aware Software System

In the testing literature, it has long been established that coverage of the whole test input space cannot be achieved, and this problem is worsened in context-aware software systems. Traditional testing techniques cannot cope with context-awareness complexities. The values of the contextual input variables and the test oracles cannot be completely anticipated during test case design. Nevertheless, they should properly reproduce the context's variability in an environment that does not impose restrictions on its variations.

The design and execution of test cases should differentiate between test inputs and context inputs, as the latter influences both the test items and the test oracles. If this abstraction is not considered when designing test cases, the software tester is left to elaborate on non-context-aware testing methods. Therefore, context-aware software testing should consider the variation of context and its influence on the test oracle and the test item inputs (Figure 2).

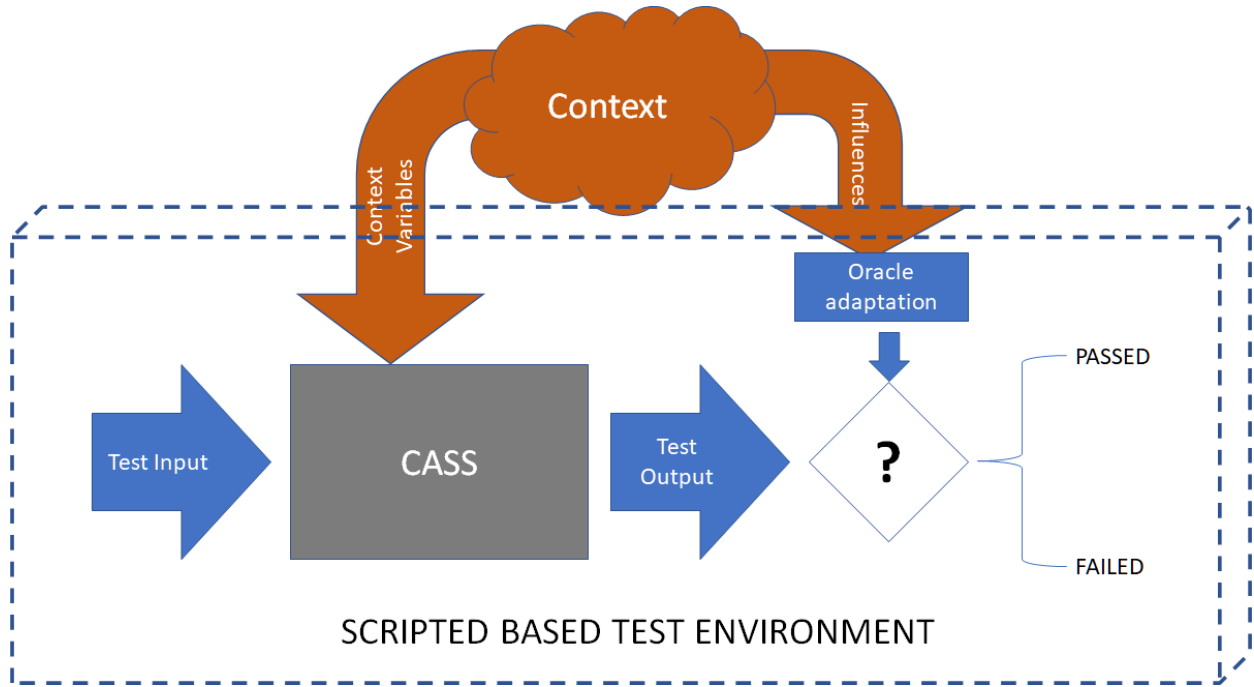

*Figure 2. A Scripted Test-Based Environment that considered context (evolved from [13])*

## 3.3 Testing Processes

Testing is a systematic process for revealing failures and, therefore, the existence of faults in software. This activity is of utmost importance for developing high-quality software systems by assessing their functional and non-functional requirements [26].

According to the ISO/IEC/IEEE 29119-2-2013 [27], the testing activities that may be performed during the test process are clustered into three main groups, i.e., *Organizational Test Process*, *Test Management Processes*, and *Dynamic Test Processes*.
In this paper, we focus on Test Management and dynamic testing processes. The Organization Test process deals with organizational policies for testing. We assume that it is not likely that the organizations will mention their policies in the literature. Because, among other issues, they would be disclosing sensitive information with potential competitors. The Management Test Processes refer to *"processes covering the management of testing for a whole test project or any test phase (e.g., system testing) or test type (e.g., performance testing) within a test project (e.g., project test management, system test management, performance test management)."* They rely on the execution of subprocesses (1) for *planning* the resources needed to execute the test activities, (2) for *monitoring* and *controlling* the execution of the planned activities, and (3) for *completing* the test process that is reached when the agreement that the testing activities are complete has been obtained. The Dynamic Test Processes *"are used to carry out dynamic testing within a particular phase of testing (e.g., unit, integration, system, and acceptance) or type of testing (e.g., performance testing, security testing, usability testing)."* They consist of four subprocesses. The *Test Design & Implementation Process* is performed to derive test cases and test procedures by exploiting combinations of testing techniques. The *Test Environment Set-Up & Maintenance Process* is used to build and maintain the running environment in which tests are executed. The test environment requirements are initially defined in the planning process, but its detailed composition normally only becomes clear once the Test Design & Implementation Process starts. In the *Test Execution Process,* the test procedures previously implemented are run in the testing environment. Finally, the *Test Incident Reporting Process* is used to report test incidents such as anomalies, bugs, defects, errors, and issues. More precisely, this process identifies and reports test failures, i.e., instances where something unusual or unexpected occurred during test execution.

The ISO/IEC/IEEE 29119-1-2013 [28] also classifies the testing techniques in static testing and dynamic testing. On the one hand, static testing is typically exploited to identify apparent defects ("issues") in documentary test items or anomalies in source code. It includes various activities, such as static code analysis, cross-document traceability analysis, and reviews. On the other hand, dynamic testing aims to reveal failures by performing executable test items with the software system. Therefore, dynamic testing's main goal is to derive test cases that must be executed on a running test item. We will point our interest only in dynamic testing techniques from now on in this paper, and we will use the terms testing technique and dynamic testing technique without distinctions. In practice, software testers usually apply one or more test design techniques for deriving test cases and procedures with the main goal of achieving a given test completion criteria, typically described in terms of test coverage measures [27], and so, to reveal as many failures as possible [29].

# 4 Research Method - RAPID REVIEW

This section summarizes the Rapid Review research protocol we have employed as the research method to guide our work. A Rapid Review (RR) aims to provide evidence on a problem at a much lower cost than a Systematic Literature Review [12]. As mentioned previously, this research effort aims to compare the advances in testing context-aware contemporary software systems since 2017, when we published two *quasi*-systematic literature reviews in this field [9,13]. At that time, there was evidence of a lack of testing techniques regarding the full validation of context-awareness. Therefore, we have developed the RR protocol with the following assumptions:
   a. CACSS have spread and are more pervasive than they were three to five years ago.
   b. The industry has had time to adopt (or develop new) techniques to deal with the context and its effects when testing software systems.

This section summarizes the RR research protocol. Its full description is available in [30].

## 4.1 Goal and Research Questions

Our RR goal is *to understand how the industry deals with the variation of context when testing Context-Aware Contemporary Software Systems.* Our goal is to characterize the software technologies that enable us to consider the variation of context in the CACSS testing processes adopted in the industry. More precisely, we are interested in distinguishing whether (and how) the industry considers the context in testing, understanding whether the context is instantiated and varied during the execution of the test cases, and whether the software system under test interacts at runtime with the running context.

To reach this goal, we defined the following three research questions:

**RQ1:** Which are software technologies supporting the **test management processes** dealing with the context for CACSS?
*Rationale*: To understand how the context and its variations are being considered when managing testing activities.

**RQ2:** Which are the software technologies supporting the **dynamic testing processes** dealing with the context for CACSS?
*Rationale*: To understand the support for executing CACSS testing. We want to understand the given support to software practitioners regarding the context and its variation during the execution of CACSS testing.

**RQ3:** How mature are the identified solutions for widespread adoption?
*Rationale*: To understand the effort that a software tester practitioner looking for testing CACSS would be required to invest in adopting one of the proposed solutions.

## 4.2 The primary studies selection process

This section provides an overview of the process we followed in selecting primary studies in this study. As Figure 2 shows, the process relies on the execution of three steps, i.e., *Automated Search Strategy*, *Primary studies selection*, and *Snowballing search strategy*. Two iterations of this process were executed (see section 5).

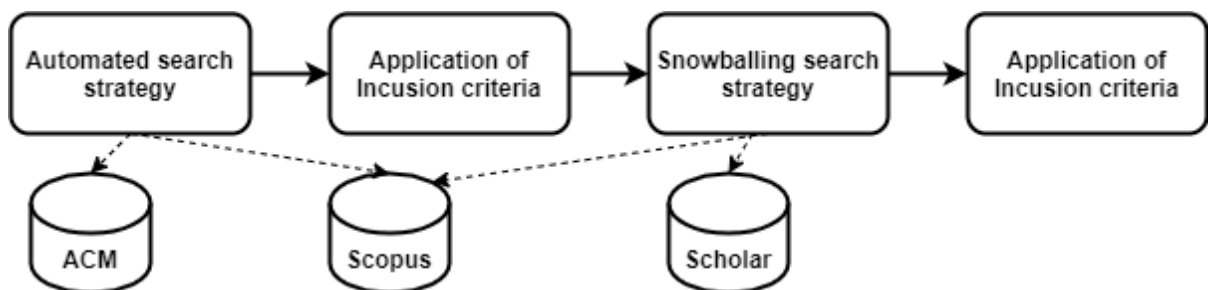

*Figure 3. Primary studies selection process*

### 4.2.1 Automatic search strategy

Scopus was the main search engine and the ACM Digital Library the secondary. We applied our search string and selection procedures in SCOPUS and repeated the exercise in ACM Digital Library. Based on this, the search string was adapted to fulfill the requirements of both search engines. The variations are built on a canonical search string developed following the PICOC method with five levels of filtering:

**Population** - contemporary software systems

Synonyms: ("ambient intelligence" OR "assisted living" OR "multiagent systems" OR "systems of systems" OR "internet of things" OR "cyber physical systems" OR "autonomous systems" OR "autonomic computing" OR "multi-agent systems" OR "pervasive computing" OR "mobile computing" OR "distributed systems" OR "cooperative robotics" OR "adaptive systems" OR "industry 4.0" OR "fourth industrial revolution" OR "web of things" OR "internet of everything" OR "contemporary software systems" OR "smart manufacturing" OR digitalization OR digitization OR "digital transformation" OR "smart cit*" OR "smart building" OR "smart health" OR "smart environment")

**Intervention** - Software Testing

Synonyms: ("test* management" OR "test* planning" OR "test* monitoring" OR "test* control" OR "test* completion" OR "test* design" OR "test* type" OR "test* implementation" OR "test* environment" OR "test* execution" OR "test* reporting" OR "software test*" OR "software validation" OR "software verification")

**Comparison** – no

**Outcome** - Software Testing Technologies

Synonyms - ("technique" OR "technolog*" OR "method" OR "activity" OR "tool" OR "process" OR "practice" OR "mechanism" OR "instrument" OR "task" OR "service" OR "strategy")

**Context** – Context Variation

Synonyms - "variation" OR "context" OR "context awareness" OR "context variation"

### 4.2.2 Selection procedure

In each of the two iterations, the selection process (see figure 2) was decomposed into the following activities described in this section. The JabRef Tool[1] has been used to manage and support the selection procedure. In each of the activities, we voted for the source's inclusion and discussed the motivations for conflicting votes (see section 5.1).

1. Automated search process. We used the automated search process as the main strategy.
    a. Run the search string in Scopus.
        i. Apply the inclusion criteria based on the paper Title;
        ii. Apply the inclusion criteria based on the paper Abstract;
        iii. Apply the inclusion criteria based on the paper Full Text, and;
    b. Run the ACM search string in ACM.
        i. Apply the inclusion criteria based on the paper Title;
        ii. Apply the inclusion criteria based on the paper Abstract;
        iii. Apply the inclusion criteria based on the paper Full Text, and;
2. Snowballing search process.
        i. Execute backward (one level) and forward snowballing (using Google Scholar and Scopus):
            1. Apply the inclusion criteria based on the paper Title;
            2. Apply the inclusion criteria based on the paper Abstract;

---

[1] http://www.jabref.org/

3. Apply the inclusion criteria based on the paper Full Text.

### 4.2.3 Inclusion criteria

The following inclusion criteria were defined:

1. The paper must be in the context of **software engineering**; and
2. The paper must be in the context of **context-aware contemporary software systems**; and
3. The paper must report a **primary study**; and
    a. The **primary** study should be conducted within an **industrial setting**, and**;**
    b. The paper must report an **evidence-based study** grounded in empirical methods (e.g., interviews, surveys, case studies, formal experiments, and others).
4. The paper must provide data to **answering** at least one of the RR research questions by showing that dealing with context during testing was a concern, and
5. The paper must be written in the **English language**.

By default, all papers that do not fulfill all the previous criteria were excluded. Duplicates were also excluded. Furthermore, we are only interested in studying peer-reviewed sources.
We developed a shared understanding among the team regarding the concepts included in these criteria during our protocol execution. We develop this further in section 5.1.

### 4.2.4 Snowballing search strategy

Forward and backward snowballing [31] was used as a complementary search strategy. We seeded the snowballing process with the selected sources selected after applying the selection procedure (see section 4.2.2) to the sources identified from the automated search process.

Regarding the initial seeds, as per the guidelines, we reviewed and evaluated the influences that those sources could excerpt on the snowballing process [31]. Therefore, we decided to include all sources, as no research group or author seemed to gravitate toward the initial seeds. Furthermore, in the second iteration (see section 5), the seed sources were complemented with sources suggested by colleagues who reviewed the protocol and the first round of results.

## 4.3 Data extraction process

We designed a data extraction form to extract the information needed to answer the research questions. This form's pertinence became clear since it fine-combed the sources for their suitability to answer the research questions. The following fields were included in the data extraction form:

- Description: We looked for a high-level summary, in the author's word, of the proposal.
- Study type: A description of the type of experimental study that was described in the source.
- Application domain: A description of the industrial setting in which the study took place.
- Type of software system: A description of the possible types of systems. We expected to identify things like unmanned flight, self-driving cars, IoT systems, and others.

In addition to the previous fields, we included one field for each of the research questions. Excerpts that directly answered the research questions were to be extracted in these fields. Finally, during the results' reporting, we added a follow-up table to extract complementary evidence for writing the summaries in section 6.1 on consistent paragraphs followed by a field where the paper presented evidence to answer the research question (see section 6.2).

# 5 Execution of the protocol

The research protocol was defined and firstly executed in 2019 and updated in the second half of 2020 (see figure 3). We reported our initial execution in a technical report [30] and took feedback from the community. We considered the feedback and re-executed our protocol to update the sources and improve the inclusion criteria application's calibration. The main two criteria that filtered out the sources are whether the source presented a CACSS and whether the source empirical setting was an industrial application (see section 5.1). These criteria taken together narrowed the amount of available research significantly in our area of interest. Figure 4 summarizes the overall search process.

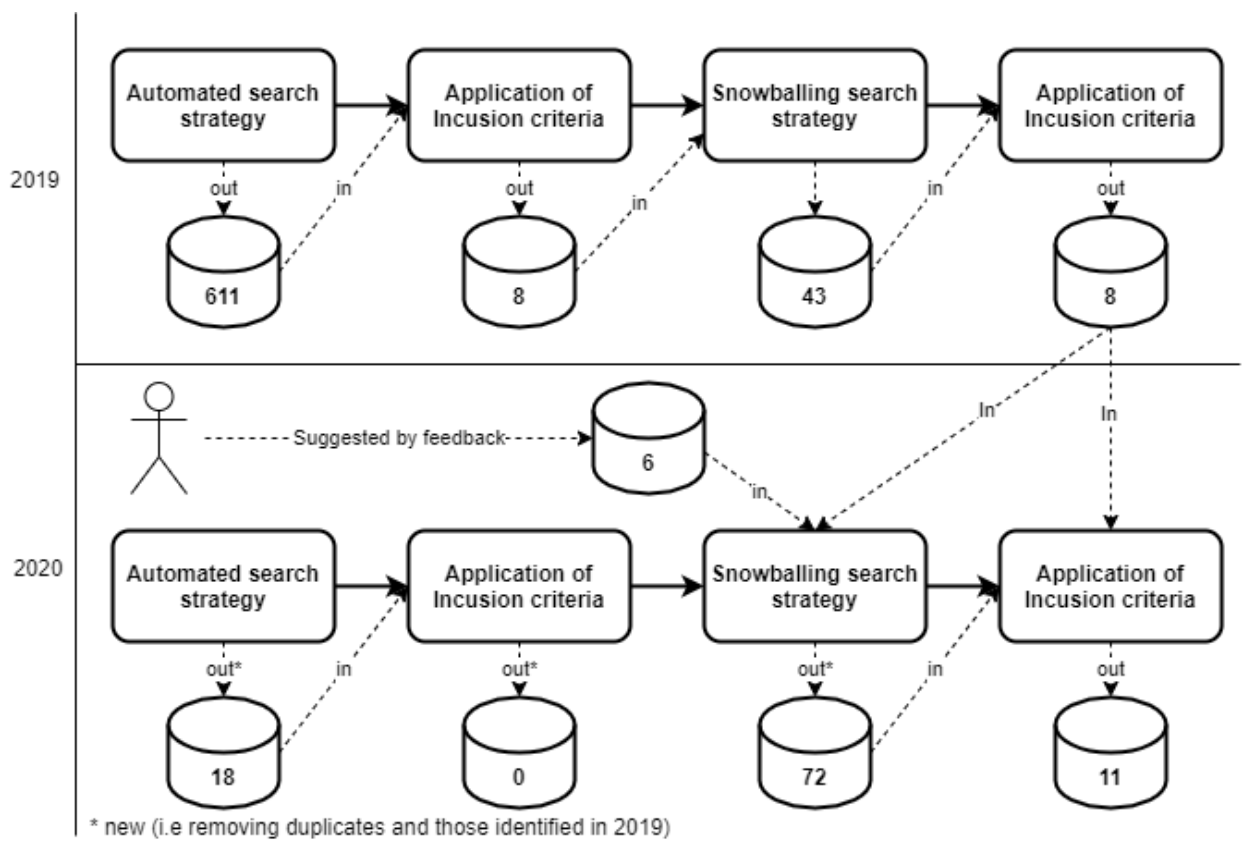

Figure 4. Rapid Review Execution Process

## 5.1 Achieving consistency with the inclusion criteria

We invested much effort in establishing the criteria' consistency among the different researchers when applying the inclusion criteria at the different RR process stages. In all its stages, the selection was adopted by consensus. This section presents a few examples of the research team's discussions to convey our rationale regarding the inclusion/exclusion criteria.

Regarding criterion 2, we extensively discussed what qualifies a *context-aware contemporary software system* (see Section 3.1). Upon reflection, most of these discussions revolved around the complexity of the software system under test. Our decision draws from Lewis's conceptual pragmatism [32]. Thereby, we probed the test item described in the candidate sources to determine whether the added complexities needed for considering the context and its variation were worth making to test such software. This exercise led to the rejection of several sources. The proposal was evaluated in software under test that was indeed a CSS (even if it were within an industrial setting) to merit the concepts' application. For instance, in [33], the proposal considers context-awareness. It conveys how the software under test is expected to be deployed in an industrial setting. However, the complexities of the defined setting for the evaluation are too simple for our purposes. We make explicit here that we are not making a judgment call on the pertinence of that system to the research goals expressed in [33]. We are making a judgment call on the pertinence of that software under test to our study's goals. Furthermore, we note to the reader that this judgment was made after going through the data extraction process for this source, conveying how suitable the decisions we are taking care of in the selection process.

Regarding criterion four, we want to clarify how we have interpreted that the primary study was conducted within an **industrial setting.** Our goal was to identify evidence of testing contemporary software systems in **industrial settings**. Therefore, our criterion involved that the source must make a clear statement that the Test Item was intended to be deployed to the production environment. We wanted to weed out Test Items developed for the research, even if these were based on real-life applications and/or with an industrial sponsor's support. For instance, [34] was evaluated in all process steps of our RR, passing the inclusion criteria until the Data Extraction process. At this point, we noticed that the authors commented that they "implemented the SVS (Smart Vacuum System) application as a completely autonomous robot that receives input from sensors and responds accordingly." [34]. As a result, we agreed that these implementation types could not be judged to belong to an industrial setting. There is an explicit statement that the researchers implemented the Test Item. Interestingly, several other sources suffered the same analysis, [34], [35,36] all contain similar statements that can only be identified in a thorough reading (or during the Data Extraction process).

On the other end, [37] presents an interesting case. The proposal is relevant to this research, but during the data extraction, it became evident that only the second use case described in the paper presents evidence of an industrial setting.

Regarding inclusion criterion 6, three conditions are included: the paper must address our phenomenon of interest, the system under test must be context-aware, and the proposal must show evidence of SUT testing. For instance, [33,38] realized there was no evidence of the test cases' execution upon data extraction.

## 5.2 Enacting the data extraction form

The data extraction form in this section is presented as an example, and for brevity, only excerpts of the full data extraction form are demonstrated. We intend to convey our process and how we enacted the guidelines set out by our research protocol (see Section 4.5). Table 1 presents the data extraction form for S1[37].

| Field | Example Extraction excerpt |
|---|---|
| Abstract | <A full text of the abstract was extracted. We mainly used this as a reference for discussions. For brevity, we do not include it here> |
| Description | "in this article, we present an approach, named context-based multiinvariant detection (CoMID), to automatically generating invariants for specifying developers' implicit assumptions and checking these invariants for detecting when a cyber-physical program has entered an abnormal state at its runtime. CoMID addresses the preceding challenges with its two techniques: context-based trace grouping and multiinvariant detection." [37] |
| Study type | "We present the evaluation of our CoMID approach, including comparing it with two existing approaches. We select three real-world cyber-physical programs, namely, NAO robot .., and six-rotor UAV …. as the evaluation subjects." [37] |
| Type of Modern software system | <Authors make no explicit statement as to this field. For our purposes, we can abstract from the 'Study type' field and left this field empty> |
| **Answer to research questions** | |
| RQ1 | <No data extracted for this source> |
| RQ2 | "CoMID works in four steps, which are as follows: 1) it first executes program P in the environment E to collect safe execution traces, i.e., no failure condition triggered (Step 1:trace collection); … We implement CoMID as a prototype tool in Java 8" [37] |
| RQ3 | <No data extracted for this source> |
| Summary paragraph | |
| Summary | "To address these challenges, we in this article present an approach, named Context-based Multiple Invariants Detection (CoMID), to automatically generating invariants and using them for effective detection of abnormal states for cyber-physical programs." [37] |
| How context is taken into account | "...For the program context, CoMID records the statements executed in each iteration through program instrumentation. For the environmental context, CoMID records values of environmental attributes using their tough system call at the beginning of each iteration (i.e., once CoMID recognizes a new iteration)...." [37] |

| Description of the empirical evaluation | "We experimentally evaluate our CoMID approach on three real-world cyber-physical programs: a 4-rotor unmanned aerial vehicle (4-UAV), a 6-rotor unmanned aerial vehicle (6-UAV), and an NAO humanoid robot….." [37] |
|---|---|
| Test Item description | "CYBER-PHYSICAL software programs (in short as cyber-physical programs) integrate cyber and physical space to provide context-aware adaptive functionalities. An important class of cyber-physical programs is those that iteratively interact with their environments. Examples of such programs are those running on robot cars, unmanned aerial vehicles(UAVs), and humanoid robots." [37] |
| Technique description | "CoMID addresses the preceding challenges with its two techniques: context-based trace grouping and multi invariant detection. <br> 1)Context-based trace grouping:….. <br> 2) Multi Invariant detection:...." [37] |
| Environment Description | "In the experiments, all invariants should be generated based on the execution traces collected from the selected experimental subjects. One execution trace is a concrete run of a subject program in a certain scenario." [37] |

*Table 1. Data extraction form for (Qin et al. 2020)*

## 5.3 Final set of selected sources

Table 2 presents the final set of sources that were selected through the execution of the Rapid Review research protocol:

| Code | Full Reference |
|---|---|
| S1 | Ma, T., Ali, S. & Yue, T. Modeling foundations for executable model-based testing of self-healing cyber-physical systems. *Softw Syst Model* 18, 2843–2873 (2019). https://doi.org/10.1007/s10270-018-00703-y [39] |
| S2 | Arrieta, A., Sagardui, G., Etxeberria, L. et al. Automatic generation of test system instances for configurable cyber-physical systems. Software Qual J 25, 1041–1083 (2017). https://doi.org/10.1007/s11219-016-9341-7 [40] |
| S3 | Seung Yeob Shin, Shiva Nejati, Mehrdad Sabetzadeh, Lionel C. Briand, and Frank Zimmer. 2018. Test case prioritization for acceptance testing of cyber-physical systems: a multi-objective search-based approach. In Proceedings of the 27th ACM SIGSOFT International Symposium on Software Testing and Analysis (ISSTA 2018). New York, NY, USA, 49–60. DOI:https://doi.org/10.1145/3213846.3213852 [41] |
| S4 | Seung Yeob Shin, Karim Chaouch, Shiva Nejati, Mehrdad Sabetzadeh, Lionel C. Briand, and Frank Zimmer. 2018. HITECS: A UML Profile and Analysis Framework for Hardware-in-the-Loop Testing of Cyber-Physical Systems. Proceedings of the 21st ACM/IEEE International Conference on Model-Driven Engineering Languages and Systems (MODELS '18). New York, NY, USA, 357–367. DOI:https://doi.org/10.1145/3239372.3239382 [42] |
| S5 | Mariam Lahami, Moez Krichen, Mohamed Jmaiel. Safe and efficient runtime testing framework applied in dynamic and distributed systems |
| | Science of Computer Programming. Volume 122,2016,Pages 1-28,ISSN 0167-6423, https://doi.org/10.1016/j.scico.2016.02.002. [36] |
| S6 | Fröhlich J., Frtunikj J., Rothbauer S., Stückjürgen C. (2016) Testing Safety Properties of Cyber-Physical Systems with Non-Intrusive Fault Injection – An Industrial Case Study. In: Skavhaug A., Guiochet J., Schoitsch E., Bitsch F. (eds) Computer Safety, Reliability, and Security. SAFECOMP 2016. Lecture Notes in Computer Science, vol 9923. Springer, Cham. https://doi.org/10.1007/978-3-319-45480-1_9 [43] |
| S7 | Raja Ben c, Annibale Panichella, Shiva Nejati, Lionel C. Briand, and Thomas Stifter. 2018. Testing autonomous cars for feature interaction failures using many-objective search. In Proceedings of the 33rd ACM/IEEE International Conference on Automated Software Engineering (ASE 2018). Association for Computing Machinery, New York, NY, USA, 143–154. DOI:https://doi.org/10.1145/3238147.3238192 [44] |
| S8 | R. Ben Abdessalem, S. Nejati, L. C. Briand and T. Stifter. Testing advanced driver assistance systems using multi-objective search and neural networks. 2016 31st IEEE/ACM International Conference on Automated Software Engineering (ASE), Singapore, 2016, pp. 63-74. DOi: https://doi.org/10.1145/2970276.2970311 [45] |
| S9 | R. Ben Abdessalem, S. Nejati, L. C. Briand and T. Stifter, Testing Vision-Based Control Systems Using Learnable Evolutionary Algorithms, 2018 IEEE/ACM 40th International Conference on Software Engineering (ICSE), Gothenburg, 2018, pp. 1016-1026, doi: https://doi.org/10.1145/3180155.3180160. [46] |
| S10 | Y. Qin, T. Xie, C. Xu, A. Astorga and J. Lu, CoMID: Context-Based Multiinvariant Detection for Monitoring Cyber-Physical Software. In IEEE Transactions on Reliability, vol. 69, no. 1, pp. 106-123, March 2020, doi: https://doi.org/10.1109/TR.2019.2933324. [37] |
| S11 | Y. Qin, C. Xu, P. Yu, J. Lu. SIT: Sampling-based interactive testing for self-adaptive apps. Journal of Systems and Software. Volume 120, 2016. Pages 70-88. ISSN 0164-1212. https://doi.org/10.1016/j.jss.2016.07.002. [47] |

*Table 2. The final set of selected papers in the Rapid Review*

# 6 Data Analysis

In this section, we report a detailed analysis of the extracted data. First, we briefly describe the testing techniques reported in the selected papers. We characterize these testing techniques and answer the research questions we proposed based on the evidence we found.

## 6.1 Summary of the included papers

This section contains a summary of the selected sources. We added this section to provide the readers with an overall understanding of the selected sources.

S1. Ma et al. [39] present a model-based approach for self-healing software systems. Their approach includes a modeling framework and an execution environment. Their approach focuses on dealing with the uncertainties bright to the software systems by perceiving the context. The authors evaluate their approach on a computer-based simulation, in which their approach is applied to nine self-healing software systems. Test cases are designed in the modeling framework and executed simultaneously in the model and the system.

S2. Arieta et al. [40] propose testing individual configurations from software product lines of cyber-physical systems. The individual configurations are selected by employing software-based simulation. Their approach models the Software product line of the CPS, and simulation is used to evaluate the interactions between the CPS and the context dynamically. The authors evaluate their approach in a computer-simulated environment. Their case study subject is a model of an Unmanned Aerial Autonomous Vehicle. Test cases are generated based on an initial set of test cases evaluated against the SPL-CPS modeled variability points.

S3. Shin et al. [41] propose a search-based approach for prioritizing acceptance test cases for CPS. Interestingly, while context variation is not explicitly addressed in this proposal, the authors consider the problems brought by uncertainties in the environment's perception. That consideration is reflected in the case study. The presented case study involves a simulation environment where software for a satellite system is evaluated. Test cases are defined in a domain-specific language, and the author's search-based proposal optimizes the selection and execution of the test cases.

S4. In another research work, Shin et al. [42] present an approach for specifying and analyzing hardware in the loop test cases for cyber-physical systems. Their approach deals with uncertainties that context can realize in the software under the test's environment. To evaluate their approach, they present a case study where the test item is a satellite system. Their main aim with the case study is to evaluate their test cases' quality concerning the approach capacity to estimate the Test Cases' execution time. Their case study environment simulates an in-orbit test setup that includes all hardware and communication protocols.

S5. Lahami et al. [36] present a runtime approach to detect and execute test cases affected by context changes. They evaluate their approach in a case study involving a healthcare application deployed within a controlled environment in their lab. The authors introduce changes to the system's components under test to observe how the approach senses the context and selects test cases from a set of initially defined test cases.

S6. Fröhlich et al. [43] present a method for evaluating the safety of critical context-awareness systems' failures. Their underlying assumption is that combination of input values from context sensing components of the software system. Their approach consists of injecting faults and observing how safely the CASS fails. They evaluate their approach in a steering system of an autonomous car within a simulated environment custom-built to their proposal.

S7. Abdessalem et al. [44] deal with conflicting feature interactions from feature composition of self-adaptive behavior driven by context. Context is considered in this proposal as it is the driver for triggering the adaptations. In this proposal, the authors reframe the problem of finding undesired feature interaction as a search problem. They evaluate their approach using three systems from a self-driving car (automated emergency braking, adaptive cruise control (ACC), and traffic sign recognition). Executable models of these systems are simulated as the search-based approach identifies conflicting feature interactions.

S8. Abdessalem et al. [23] propose an automated technique to test complex Advanced Driver Assistance Systems (ADAS) using physics-based executable models of the system and its environment. As an evaluation, they presented an exploratory case study using an ADAS called Pedestrian Detection Vision-based (PeVi) system. The PeVi was created to help drivers detect pedestrians' proximity (either human or animal) during low visibility situations. The case study focused on identifying high-risk test scenarios (the ones that are more likely to reveal critical failures) using a Multi-objective search. The context was captured as a domain model, specifying a restricted simulation environment and serving as input data (which can be either static or dynamic properties) to test scenarios. Since testing PeVi with real hardware and environment would be dangerous, time-consuming, and costly, they used physics-based simulation platforms as a testing environment.

S9. In this work, Abdessalem et al. [24] focus on simulation-based testing of vision-based control systems from the automotive domain (ADAS). A domain model is used to capture the test input space and output. The input variables were classified into two categories: Static input variables (the values still fixed during the entire ADAS simulation) and Dynamic objects (indicating objects that change their position during the simulation). As an evaluation, they presented an exploratory case study of an Automated Emergency Braking (AEB) system from the automotive domain. The objective was to evaluate the ability of a search-based testing algorithm called NSGAIIDT to investigate critical regions in ADAS input spaces to identify critical test scenarios. As a test environment, they used a widely-used and commercial ADAS simulator called PreScan simulator.

S10. Qin et al. [37] present an approach for dealing with context input states that can lead to failures. In their approach, context is left to vary as there are no imposed restrictions on context variables' values. However, they demonstrate this approach offline, as execution captures of the system under test are performed in the field for later post-processing in the lab to evaluate their approach. As mentioned in section 5.1, in this paper, we are interested in evaluating their approach to the Unmanned Autonomous Vehicle. Their main idea is to define rules for context value that, if broken, would lead to failure (they call these invariants), then observe the software system's behavior to abstract these invariants states and execute them in a model of the system to observe the behavior.

S11. Qin et al. [47] present an approach for selecting test cases for self-adaptive context-aware software systems. Context variation is considered in terms of uncertainty in the adaptation rules and uncertainty in the sensor's capacity to measure the environment accurately. They evaluate their approach with three case studies for which they implement their approach in a simulated environment. In their approach, test cases are selected by sampling input space parameters after sorting them into different categories according to the abstraction mentioned above.

Progress has been made since the secondary studies in 2015 by observing the identified evidence from the included papers summaries. Mainly because in 2015, there was no evidence of industrial applications. The second outstanding observation is that the selected papers did not present technologies to support either Organizational Test Process or Test Management Processes. As we mentioned in section 3.3, it was not expected to find evidence on Organizational Test Process. However, a lack of technologies to support the test management process was not expected. At the same time, all the selected papers presented technologies supporting Dynamic Test Processes. In the following section, we characterize these technologies and describe how they are applied in industrial settings to test CACSS.

## 6.2 Analysis of the identified papers

In the selected primary studies, we did not find any evidence regarding software technologies supporting the management of the test activities dealing with the context for CACSS, RQ1.
In the following, we describe the analysis of the extracted data that allowed (1) to characterize the technologies for supporting the testing execution (RQ2) and (2) to understand whether the identified solutions are mature enough for their widespread adoption (RQ3).

### 6.2.1 Analysis of the software technologies supporting the execution of test activities dealing with the context.

To characterize the software technologies supporting the execution of CACSS testing processes, we analyzed the extracted data to understand each proposed testing approach:
1. the type of CACSS under test (Item Under Test);
2. the goal of the testing technique (Testing Goal);
3. how the context variation has been taken into account (Context Variation);
4. the supporting testing environment (Testing Environment);
5. the adopted dynamic testing technique (Testing Technique).

The data shown in Table 3 reports information about the type of CACSS considered in the selected primary studies and the main goal of the testing technique they proposed. As for the tested items, most of the proposed techniques had been applied to Cyber-Physical and Automotive Software Systems. Table 4 outlines how the variation of context is considered in the testing techniques and how it is implemented in the testing environment. Table 5 describes the testing techniques proposed for testing CACSS. These tables' content gives the evidence to support the discussions we reported in section 6.3 to answer the research question RQ2.

| ID | Test Item | Testing Goal |
|---|---|---|

| S1 | Self-Healing Cyber-Physical System (SH-CPS). CPS can perceive that it is not operating correctly and, without human intervention, makes necessary changes to its architecture or behaviors to restore itself towards a normal state. | To test the system against an executable test model. Such comparison also reveals how likely the system's behaviors are going to deviate from the modeled behaviors. The likelihood can be used as a heuristic to find the optimal stimuli sequence to detect faults efficiently. |
|---|---|---|
| S2 | The Cyber-Physical System AR.Drone, an Unmanned Aerial Vehicle (UAV) developed by Parrot for the market of video games and home entertainment. | To automatically generate test system instances for each configuration of configurable CPSs. |
| S3 | Satellite cyber-physical system. | To improve the quality of a Hardware in the Loop (HiL) test suite based on execution time, the hardware damage risks, and the criticality of a test case (TC), i.e., how important the functionality targeted by TC is in the system under test. |
| S4 | Satellite cyber-physical system. | 1. To formally ensure that Hardware in the Loop (HiL) test cases properly manipulate and interact with the SUT and any additional instruments that provide inputs to the SUT or monitor its outputs.<br>2. To estimate, via simulation, the execution times of HiL test cases and improve HiL test planning. |
| S5 | Healthcare adaptable and distributed system. | 1. To perform safe and efficient, in terms of $i$ execution time and memory consumption, platform-independent tests.<br>2. To reveal inconsistencies, functional and non-functional faults are introduced after dynamic changes. |
| S6 | The car steering of an electric car is built using software-intensive electronic devices (a cyber-physical system). | To demonstrate the system safety in different configurations of varying degrees of redundancy. |
| S7 | A self-driving system including interaction features to automate independent driving functions. Examples of features are automated emergency braking (AEB), adaptive cruise control (ACC), and traffic sign recognition (TSR). | To generate test inputs that expose undesired feature interactions. |
| S8 | To evaluate a Pedestrian Detection Vision-based (PeVi) system (a vision-based advanced driver assistance system - ADAS). | To obtain test scenarios that stress several critical aspects of the system and the environment with a reduced execution time. |

| ID | | |
|----|---|---|
| S9 | To evaluate an Automated Emergency Braking (AEB) system (a vision-based advanced driver assistance system -ADAS). | 1. To reduce the generation time of critical test scenarios.<br>2. To characterize critical regions understandably and intuitively that may help engineers debug their systems, identify hardware changes to increase ADAS safety, and specify conditions that are likely to lead to ADAS failures. |
| S10 | To evaluate three real-world cyber-physical programs: a four-rotor unmanned aerial vehicle (4-UAV), a six-rotor unmanned aerial vehicle (6-UAV), and an NAO humanoid robot. | To automatically generate invariants and detect abnormal states for cyber-physical programs' executions. |
| S11 | Self-adaptive applications that typically are deployed in embedded systems or smartphone platforms. | To find bugs in the system under test at a reduced cost time. |

*Table 3. Item Under Test and Testing Goal*

| ID | Variation of Context | Testing Environment |
|----|---------------------|---------------------|
| S1 | The context variation is described through the uncertainties that may arise from interactions between SH-CPSs and their environments. Uncertainties are modeled in UML language. According to their uncertainty level, the values assumed at run time by the different types of uncertainties are defined in five different ways, varying from *complete certainty* to *total ignorance*. | The testing environment executes the simulators/emulators, the test model, and the system coordinately to fulfill the executable model-based testing. A specific component is needed to orchestrate the execution of diverse objects that are typically implemented by distinct modeling and/or programming languages and from the perspectives of diverse domains. For instance, the test model, modeled by state machines, exhibits discrete behaviors. On the contrary, a simulator, modeled by differential equations, presents continuous semantics. |
| S2 | The context variation is considered through a context environment model that simulates the physical world in which the CPS resides and interacts. A UML model (representing the physical world's unpredictability interacting with the CPS) describes the context environment. | It can run functions implementing the context environment model. These functions simulate the environment directly affecting the item under tests, such as the wind, the obstacles, and the information related to the person that the UAV must follow. |
| S3 | A variable assumes a special value, denoted *unknown,* when the variable's actual value can be determined only at the time of testing. The *unknown* value is used for variables that depend on uncertain environmental factors such as temperature. | It simulates test case executions and monitors these executions to keep track of how the components change states. An ad-hoc domain-specific language (DSL) is needed to perform such a simulation. Test case specifications are also based on log files from real-world executions of test cases in previous in-orbit testing campaigns performed on satellites and similar infrastructures. |
| S4 | A special *Unknown* literal represents a value that can be determined only at the time of HiL testing and is unknown at test specification. *Engineers use unknown literal* values that depend on uncertain environmental factors that are not a-priori-known, such as temperature. The Unknown literals are replaced with random-number generators, which yield random numbers with the same type as the respective uncertain variables and within ranges specified by engineers. | It is a model checking and simulation environment. Simulation is used to compute the execution time estimations for HiL test cases. Historical data files from real-world executions of the tests in previous in-orbit testing campaigns performed on satellites and similar HiL platforms are needed. |
| S5 | The context variation is expressed in terms of component dependencies formally represented using Component Dependency Graphs and Component Dependency Matrices. | It is identical to the software execution environment, same design, same implementation. |
| S6 | The context of each fault-tolerant function can be modeled as a graph where the nodes representing the context can be of three types. i.e., input from the platform periphery (sensors), data processing in the platform control computer, and output to the platform periphery (actuators). | It resembles a minimalistic system under test made by two real sensors and the central steering control application. |
| S7 | The mobile environment objects and the static environment properties (e.g., weather condition and road shapes for self-driving systems) are implemented in *PreScan*, a physics-based simulator for self-driving systems. *PreScan* relies on dynamic *Simulink* models to compute cars and pedestrians' movements and capture the environment's static properties such as weather conditions and road topology. | It relies on a simulation environment where running models of the context are executed. |
| S8 | The context is considered through *PreScan* that simulates the behavior of PeVi, the vehicle, and the pedestrian. More precisely, *PreScan* is used to execute simulations of the Simulink models of the pedestrian, the car, and the PeVi system embedded into the car. | It relies on a simulation environment where running models of the context are executed. |
| S9 | The context variation is allowed thanks to PreScan to define and execute scenarios capturing various road traffic situations and different pedestrian-to-vehicle and vehicle-to-vehicle interactions. Besides, using PreScan, it is possible to vary road-topologies, weather conditions, and infrastructures in test scenarios. | It relies on a simulation environment where running models of the context are executed. |
| S10 | The context is considered through execution traces that were preliminary collected, leaving the SUT free to evolve. | It is a real controlled physical environment, i.e., an indoor area (including random obstacles and different floor materials) or a flying space delimited by one starting point and one landing place. |
| S11 | The context variation is described through the uncertainties that may arise from interactions between the app and the environment. The uncertainty specification is defined as a set of functions, mapping a given environment's output parameter to its corresponding app's input parameters and the associated error range. | It relies on a simulation environment where it is possible to emulate the context surrounding the item under test. |

*Table 4. Context variation and testing environment*

| ID | Testing Technique |
|---|---|
| S1 | Executable model-based testing directly tests a system against an executable test model by executing the model and the system together, sending them the same stimuli, and comparing their consequent states. Based on the executable test model, a testing strategy is used to select the runtime stimuli to drive the execution. In this case, a random testing strategy is adopted for randomly selecting an outgoing transition to generate a testing stimulus. |
| S2 | It combines requirement-based testing and model-based testing techniques. The technique needs as input a variability management file (manually developed by test and system engineers), a CPSUT model (developed by system engineers), a model related to the context in which the CPS operates (the test engineers usually build it), a configuration file (it relates to a specific system variant that must be tested), a set of test cases (it can be either automatically generated from a behavioral SUT model or manually specified by test engineers), and a set of requirement monitors (a test oracle to verify that each requirement meets the specifications provided by the test engineers). |
| S3 | A multi-objective genetic approach to prioritize the acceptance test cases. |
| S4 | It combines formal verification and simulation-based methods. A textual language named Hardware-In-the-loop TEst Case Specification (HITECS) is defined using the UML profile mechanism to apply these methods properly. |
| S5 | It presents a model-based testing technique relying on the Testing and Test Control Notation Version3 (TTCN-3) language. |
| S6 | A fault-based testing technique. |
| S7 | To detect undesired feature interactions by algorithmically generating many-objective genetic tests. |
| S8 | Multi-objective search-based technique to obtain test scenarios that stress several critical aspects of the system and the environment simultaneously. |
| S9 | It combines evolutionary search algorithms and decision tree classification models. The classification models guide the search-based generation of tests faster towards critical test scenarios. The search algorithms refine the classification models to characterize critical regions accurately. |
| S10 | It describes a context-based multi-invariant based detection technique that automatically generates invariants for specifying developers' implicit assumptions. The technique also checks these invariants for detecting when a cyber-physical program has entered an abnormal state at its runtime. |
| S11 | A model-based technique that captures the dynamic interactions of an app under test with its environment. The model is also exploited in an adaptive sampling technique that explores the app's input space to generate test cases for testing the app's behavior. |

*Table 5. The proposed testing technique*

### 6.2.2 Analysis of how the identified solutions are ready for widespread adoption.

To understand how and if the proposed solutions are ready for being widely adopted, we evaluated their technology readiness levels (TRLs). The technology readiness levels provide a useful model for conveying the maturity of technologies. The classification was initially defined for the Apollo missions and has been extensively adopted in Europe [48]. The TRL levels are a discrete scale of nine levels from Basic research (TRL1) to Operational Readiness (TRL9). To estimate the TRL level, we used an estimator tool developed by the University of San Diego [49].

To support an evidence-based input to the estimator, we classified the sources according to the following categories of three orthogonal dimensions (see [Table 6](#)).

The column "Engineering Cycle" is based on [50], in which the author describes an Engineering cycle for technology development projects and classifies the aim of experimental work. We used this cycle to provide a uniform language to address the primary sources' experimental evaluations under study.

The column "Study Type" is based on the classification provided by [51]. This classification indicates the research method described in the primary source. We used this classification to provide uniform language and convey the capacity to generalize the sample's research results.

The column "Environment Control" is based on the classification proposed in [52], indicating the degree of control that the experimenters introduce in the environment to observe the phenomenon under study. We used this classification to convey the degree of control that the experimenter has in the environment where the empirical evaluation is described in the source tool place.

Finally, to achieve consensus on our estimates for each of these categories, we followed a Delphi-type cycle where each of the two lead researchers would classify the paper into the available categories and explain their positions. The other authors reviewed the results.

We exemplify the reasoning concerning paper S1:
- The engineering cycle is classified as "Solution Validation," as the research questions in the evaluation section aim to evaluate whether the proposed technology can achieve its intended goals. For instance, in the S1, its first research question is "RQ1: Can (the proposal) cover all relevant concepts and their relationships identified in selected self-healing cyber-physical systems?"
- Study type is classified as a "Confirmatory case study," as the research questions are framed with a positive bias towards achieving their research goal. As per the same example, RQ1 is not aimed at characterizing the proposal's capabilities but rather at confirming that the author's proposal's constructs contribute to solving the research goals.
- Environmental control is classified as "*In-Vitro*," as the paper's experimental section shows that the researchers have complete control of the environment.
- Due to all previous issues, the TRL using the estimator mentioned above tool [49], at Level 4.

[Table 6](#) shows that the proposals were classified within the design half of the engineering cycle. It means authors are still mainly concerned about the purpose fitness of the proposed solution. Furthermore, [Table 6](#) sources apply case study research. While this can be suitable to evaluate the proposals, it also limits the transferability to other application domains. We also note that the researchers had some degree of control over the environment (i.e., we could not classify any of

the proposals within the "*in-vivo*" category). That last observation is consistent that all estimations of the TRL levels are within levels 3 to 6.

| ID | Engineering Cycle | Study Type | Environment Control | Estimated TRL |
|---|---|---|---|---|
| S1 | Solution validation | Confirmatory case study | *In-vitro* | L4 |
| S2 | Solution validation | Exploratory case study | *In-vitro* | L3 |
| S3 | Solution design | Confirmatory case study | *In-virtuo* | L4 |
| S4 | Solution validation | Confirmatory case study | *In-virtuo* | L4 |
| S5 | Solution validation | Confirmatory case study | *In-vivo* | L5 |
| S6 | Solution design | Exploratory case study | *In-vitro* | L3 |
| S7 | Solution validation | Confirmatory case study | *In-virtuo* | L5-6 |
| S8 | Solution validation | Exploratory case study | *In-silico* | L3 |
| S9 | Solution validation | Exploratory case study | *In-silico* | L3 |
| S10 | Solution validation | Exploratory case study | *In-Vitro* | L4 |
| S11 | Solution design | Exploratory case study | *In-virtuo* | L5-6 |

*Table 6. Level estimators of Testing technology readiness*

## 6.3 Answering the RQs

The RR uncovered evidence on test proposals for dealing with the complexity considering the context for CACSS. Despite the breadth of the systems types for which these proposals were identified, they share the common trait that they are dealing with the complexities brought by considering the context in the testing process.

**RQ 1**: Which are software technologies supporting the test management processes dealing with the context for CACSS?

As we mentioned in section [6.1,](#) we did not identify any solution dealing with test activities management.

**RQ 2**: Which software technologies support the dynamic testing processes dealing with the context for CACSS?

The reviewed solutions focus on enabling the execution (dynamic testing) of context-aware software systems. As shown in section 5.3, technologies can be characterized from different perspectives:

1. Interesting results were obtained regarding the testing goals. We observed that these goals are mainly two. First, there are testing techniques to assess the CACSS quality by testing functional or non-functional requirements such as reliability and security (S1, S2, S5, S6, S7, S8, S10, and S11). Moreover, there are testing techniques (S3, S4, and S9) for assessing or improving a quality attribute of test suites already designed, such as performance (reducing the testing time) or reliability (excluding test cases that could damage the hardware under test).
2. The variation of context is mainly considered utilizing models that can also be executed. Usually, these models are implemented in Matlab/Simulink (S7, S8, and S9), UML diagrams, or UML metamodels (S1, S2, S4, S5, S6, and S11). *Ad-hoc* solutions relying on a DSL (S3) and source code invariants (S10) were also proposed.
3. As for the testing environment, we observed that it is mainly a simulation environment where the executable models can be executed with the software system under test (S1, S2, S3, S4, S7, S8, S9, and S11). Another possible solution is to build a testing environment like the real environment (S5, S6, and S10).
4. Regarding the adopted dynamic testing techniques, we observed that model-based (S1, S2, S5, and S11) and search-based or genetic (S3, S7, S8, and S9) approaches are the most exploited. Model-based is mainly adopted for testing the software system's quality under test. In contrast, search-based or genetic is used to improve a test suite's quality attribute or generate test scenarios. Other types of approaches had also been proposed, such as the use of formal methods (S4), fault injection (S6), and multi-invariant based (S10).

**RQ 3:** How mature are the identified solutions for widespread adoption?

Observing the analysis presented in section 6.2.2, we claim that none of these proposals are mature enough for widespread adoption. At TRL6, the proposals can only be expected to have been demonstrated in a relevant, not completely stressful environment, and transfer to another environment can come at a high cost. We have observed that all proposals require significant investment in technological development. Furthermore, though the path set by these tools can potentially be implemented in other working domains, they are still shy of higher readiness levels where the technology can be transferred to other software testers practitioners without significant re-engineering or tailoring.

# 7 Threats to validity

This section discusses the threats to our research work's validity using the categories described in [53].

Regarding internal validity, the threat to missing literature is common to all secondary studies. Nonetheless, we want to bring two points to the reader's attention. Throughout section 4, we intended to convey the thoroughness of our approach to searching the available literature. The aim of section 5.1 is to convey the criteria with which the literature was analyzed and ultimately selected. Other research works to address the challenge of testing CACSS. However, our inclusion and inclusion criteria did not pass muster to be included in this study sample.
Another point to make is that we decided to only look at white literature. Given that our working assumption is that CACSS are mainstream, it is likely that big players[2] have found ways to test these types of systems. However, it is just as revealing that - if the knowledge is available - it has not been contributed to the academic literature.

Regarding construct validity, the reader can question our construct's objectivity since it is biased by our previous research. We would claim that Rapid Reviews are designed to be executed by experts in the field. Therefore, though our approach towards the new evidence is objective, our previous knowledge informed our judgment. We would argue that this strengthens the results. Furthermore, we have taken care to expose these assumptions (sections 3.1 and 3.2), so the reader can make their judgment.

We stress that we are taking an awfully specific understanding of the constructs mentioned in our inclusion criteria regarding external validity. We would claim that we have fine-combed the literature and have made extensive discussions (section 5.1) before deciding if a source complies with our understanding of *Context Awareness*, *Testing*, *Contemporary software systems*, and *industrial setting*. We assume that this point of view strengthens the construct validity at the expense of our external validity.

Regarding conclusion validity, we have tried to provide evidence for each of our research questions' answers (section 6.3). Our analysis section fairly portrays the uncovered evidence, which is, in turn, based on a thorough and verbatim data extraction process. We believe we have provided end-to-end traceability from the sources to our answers. In addition to this, we have made a clear distinction in this paper to differentiate evidence-based answers to the research questions (section [6.3](#)) from our interpretation and discussions (section [8](#)).

---

[2] A few come quickly to mind: Tesla; Google; NASA, DJI, among others.

# 8 Discussions

## 8.1 Observations and lessons learned

Considering the RR results, our previous results, and our experience testing CACSS, we put forward some lessons learned.

1. Testing context-awareness features of contemporary software systems require a model capable of modeling the system's dynamic behavior.

   A model is a reasonable representation of the system that can be used as a surrogate of the actual system to design or improve Test Suites.

   The one pervasive theme that stems from our observation of testing CACSS is that most proposals use models to explore the system. We would claim that systems models that cannot evolve with the software system are not good for modeling the dynamic nature. These models successfully design stable context states but fail to provide insight into the system's behavior during a state transition. For instance, these models cannot perceive the variation of context that has not been defined into a set of valid states.
   As a result of this observation, we favor proposals modeling the systems' dynamic behavior as these are better suited to capture the dynamic nature of the context. Nonetheless, we have also observed proposals that, while developing dynamic models, these are used to fix values for context variables, thereby limiting the models' capacity to reproduce the varying nature of the context.

2. Technologies for developing CACSS models.
   Another issue regarding models relates to the technologies used in their development. From formal mathematical-based models to Matlab Simulink to UML-based models, we have observed various technologies. They all convey a key requirement that these models must be executable - as we will discuss shortly in the following point - models are mainly used to simulate the software system execution to derive or improve a Test Suite. In section [8.2](#), we discuss the appropriateness of these technologies for testing CACSS.

3. Testing CACSS is cost-intensive and requires the exploitation of computational resources. We would conjecture that the effort required to test CACSS is still an order of magnitude greater than the effort required to develop them. Much like we had observed in [13], the identified solutions in this RR all require significant design and development effort and, when deployed, they tend to consume significant computational resources. All solutions identified in the RR cannot be transferred to other contemporary software systems. Therefore, while the alternatives presented in this paper set a path to test CACSS, software testers practitioners are bound to reproduce the steps and technologies. Testing CACSS is more cost-intensive than testing traditional software and requires the exploitation of cost-effective techniques that reduce the (1) testing time and (2) risks of damaging expensive hardware. We learned that multi-objective, search-based, and model checking-based testing techniques could be successfully applied to reduce CACSS dynamic testing processes' costs.

4. Requirements for driving the variation of context in Test Environments vary according to the Software Development Life Cycle (SDLC). We understood in this RR that to properly reproducing the variation of context during the dynamic execution of testing processes, it is necessary to design, implement and deploy a not trivial (and perhaps not available yet) test environment. The way the dynamic context variation is provided may depend on the specific stage of the SDLC. We observed that two main solutions might be applied. In the first stages of the development life cycle, the testing environment could provide a context-aware simulated environment where the context models can drive the simulation. Towards the SDLC final stages, the testing environment should be able to resemble as much as possible the real execution environment where the CACSS will run. Possible solutions to guarantee the testing environment's context-awareness are using emulators for mocking up real devices or using the same hardware/software components composing the final execution environment.

## 8.2 Future research directions on CACSS testing

Drawing from the previous observations, we propose research directions for testing context-aware properties of contemporary software systems.

1. Evaluating the efficacy of models to represent the real world. George box's quote is often repeated: "All models are wrong, but some are useful." For testing CACSS, we are concerned with the model's capacity to represent their production environments.
2. Measuring the coverage of the Test Suites. Coverage measurements of the Test suite were not identified in this RR. From our previous results [13], we had identified an effort concerning coverage measurement [54,55]. So, we were somehow surprised the coverage measurement was not observed in this sample of papers. We would call for more research on the coverage measurement of the Test Suites regarding CACSS.
3. Functional suitability. Related to coverage is the question of generating Test Suites for evaluating the correct behavior of CCASS in the face of the variation of context. This problem is related to the sheer volume of the test cases and the capacity to solve the Test Oracle problem [56] for each possible context adaptation. Applications of metamorphic testing [57] might be a possible way towards it. This time, we were not that surprised that the primary sources did not capture this approach's applications, given the inherent complexity of applying it in industrial settings. In any case, we claim this is a research line that needs further effort and investigation.

Therefore, our preferred approach is to keep a degree of human factors in the loop [58,59] and then exploit it with simulations.
4. Safe failure of safety-critical CACSS. When looking at the reliability quality attribute, existing alternatives to testing CACSS tend to make extensive use of models and simulations to identify the combinations of inputs that might reveal a failure. This information can be captured in a Test Case and feedback into the development process for safe failing. However, as complete coverage of the test input space cannot be warrantied, eventually, the software will fail. We would suggest that a research line should be to warranty the correct safe-failing behavior of CACSS. It is a reduced problem from the Test Oracle, as it is only looking at ensuring a safe behavior of the CACSS when the set of inputs makes it fail.
5. Management of test activities for CACSS. The results of this RR raise the attention at the management of test activities. We would assume that practices and procedures should evolve to manage the context and Test Suites in the different environments that the CACSS must be dynamically tested.
6. Artificial Intelligence (AI) approaches for dealing with the complexity of the context. The RR indicated that novel testing techniques based on genetic algorithms or multi-objective search-based approaches had been successfully adopted for reducing the costs of dynamic testing processes for CACSS. We believe that this trend may lead to the use of AI in testing processes for CACSS. AI could be exploited for implementing solutions aiming to emulate the variation of context in a more realistic way as possible.

### 8.2.1 Recommendations for Testing CACSS

Finally, we wish to close this discussion section with a set of guidelines that software testers practitioners can apply when faced with the challenge of testing CACSS.

Good practice conveys that testing must be carried out in an environment as close-as-possible to the intended production environment. Each deviation from this heuristic means the test process introduces the risk that the Test Environment's behavior is different from the Production Environment. Testing is and has always been, from its definition, an exercise in risk-taking. Since its initial formalizations, testing and test case selection are driven by the test analyst, striking a balance between comprehensiveness and risk of defects being carried into production. In modern CACSS, these forces are pushed to levels to which our Test Design techniques have yet to adapt. The sheer possibilities of variables and corresponding values that context can attain make the conundrum of selecting a suitable set of test cases even more challenging. Worsened still by the cost of reproducing the production environment for testing.

We put forward three propositions that must be taken into consideration for developing technologies to test CACSS:

1. Conceptual. Accept the nature of context and differentiate that the test item is subjected to different input types - The *test input and the input from the context*. First, the context in which the Test Item is being executed will vary through external forces and by the interaction with the test item. Test Management and Test Design Techniques for testing CACSS must accept this. Secondly, test input and inputs from the context are two different types of inputs. The test input can and should be planned by the test analyst— the test input results from applying a Test Design Technique during a Test Case definition. Model and simulation can be used to enhance the capacity of the Test analysis. However, as we mentioned, we prefer solutions that keep the test analyst in the loop. Context input is not within the control of the test analyst. They come from the interaction of the test item with the environment. Test Design Techniques for CACSS must accept this lack of control and design the test cases around it. We recommend that software testers accept this behavioral nature of context [60] when testing CACSS and gradually relinquish control of the context to the environment as the software system's development stabilizes through its journey through the software development life cycle.

2. Technical. Start with a dynamic system model. All successful observed experiences of testing CACSS started by modeling the software system. This model must support dynamic simulation. We have made the case that state-based simulation will not capture the nature of the context. Computing resources can then be leveraged to evaluate combinations of input values towards fulfilling a quality attribute (performance, reliability). In short, the idea is to have a system model to aid the test analyst with the development of Test Cases and explore computing simulations to minimize the risk that a combination of Test Cases can provide contextual inputs that leads to a failure (reliability) and it is fed to the system during its execution in a production environment.

3. Procedural. Manage the context - and the exposure of the test item to the context - throughout the SDLC. Testing CACSS is expensive. Therefore, there is little use in investing time, effort, and money into these complex solutions for testing CACSS if there is no reasonable assurance that the software systems behave as expected and have been developed following accepted quality guidelines. We claim that other elements of the testing lifecycle must be evolved to accept CACSS. As far as we could investigate, we have already noticed that our research did not find any study that caters to the influence on the context of the "Organizational Test Process" or the "Test Management Process."

# 9 Conclusions

Testing context-aware contemporary software systems are challenging, mainly because of the behavioral nature of the context. The variation of context in the environment during the software system's lifecycle generates an exponential increase in the test input space. This increase worsens the challenge for selecting suitable Test Suites to guarantee the test item's quality.

We have been working on the testing of CACSS for more than half a decade. This paper presents the result of a rapid review aimed to identify alternatives to testing CACSS that have been observed and evaluated in industrial settings. We commissioned this RR under the assumption that CACSS are mainstream (from Self-driving autonomous vehicles to systems of systems that source data from multiple IoT sources). Therefore, they are being tested in the industry.

Our results show that the focus of current research initiatives is aimed at generating or improving Test Suites that can deal with the variation of context and the sheer volume of test input possibilities. To achieve this, we have observed strategies that revolve around the same two concepts. First, creating a dynamic model of the system. Second, using computer-based simulations to identify or improve Test Suites. Besides, we observed that all of the identified solutions are at a relatively early development stage and are domain-specific. It means that there is still no technology that can be readily transferable for testing CACSS. Finally, we observed that none of the technologies deals with managing the testing process throughout the SDLC.

In addition to our results, we present a critical reflection on the state of technology art and practice of testing CACSS. Regarding the state of practice, we make two important claims to help software testers practitioners. First, that testing CACSS is effort-intensive and costly in terms of computational resources. Second, that testing CACSS starts with having a useful model of the CACSS. We conceptualized these claims into three propositions that software testers practitioners must abide by when developing their technologies for testing CACSS. Conceptual - *Accept the nature of context and differentiate that the test item is subjected to different input types: the test input and the input from the context.* Technical - *start with a dynamic model of the system.* Procedural - *Manage the context - and the test item's exposure to the context - throughout the SDLC.*

Finally, we set out a few research lines for evolving state of art on testing CACSS. We argued for five problems, from managing the testing lifecycle to measuring context-aware Test Suites' coverage. We claim that current research has mostly focused on the dynamic test execution and that the research community must address other aspects of the testing lifecycle to evolve the body of knowledge regarding the testing of context-aware contemporary software systems.